\documentclass[fleqn,usenatbib, useAMS]{mnras}
\usepackage{newtxtext,newtxmath}

\usepackage[T1]{fontenc}
\usepackage{ae,aecompl}
\usepackage{graphicx}	
\usepackage{amsmath}	
\usepackage{amssymb}	
\usepackage{multirow}
\usepackage{xcolor}
\usepackage{lipsum}

\newcounter{defcounter}
\newcommand{\msun}{\mbox{\,$\rm M_{\odot}$\,}}        
\newcommand{\lsun}{\mbox{\,$\rm L_{\odot}$\,}}        

\newcommand{\Lir}{\mbox{\,$\rm L_{IR}$\,}}

\newcommand{\CI}{\mbox{\,C{\sc i}\,}}

\newcommand{\mol}{\mbox{\,$\rm{H_2}$\,}}

\newcommand{\td}{\mbox{\,$\rm{T_d}$\,}}
\newcommand{\tx}{\mbox{\,$\rm{T_{kin}(CI, LTE)}$\,}}

\newcommand{\tk}{\mbox{\,$\rm{T_k}$\,}}

\newcommand{\mic}{$\mu $m\,}
\newcommand{\cc}{\mbox{$\rm{cm^{-3}}$\,}}

\title[Subthermal excitation of the C{\sc i} lines]{The subthermal excitation of the  C{\sc i} lines in the molecular gas reservoirs of galaxies:  its significance and potential utility}

\author[Papadopoulos, Dunne, \& Maddox]{
P. Papadopoulos$^{1,2,3}$\thanks{E-mail: padelis@auth.gr}, L. Dunne,$^{2}$, \& S. Maddox,$^{2}$ \\
$^{1}$ Department of Physics, Section of Astrophysics, Astronomy and Mechanics, Aristotle University of Thessaloniki,\\ Thessaloniki, GR-54124, Greece\\
$^{2}$ School of Physics and Astronomy, Cardiff University, Queens Buildings, The Parade, Cardiff, CF24 3AA, UK \\
$^{3}$ Research Center for Astronomy, Academy of Athens,  Soranou Efesiou 4, GR-11527, Athens, Greece}

\date{Accepted 2021 October 27. Received 2021 October 22; in original form 2021 June 24}

\pubyear{2021}

\begin{document}
\label{firstpage}
\pagerange{\pageref{firstpage}--\pageref{lastpage}}
\maketitle

\begin{abstract}
  We examine a sample of 106 galaxies for which the total luminosities
  of the two fine  structure lines $^{3}P_1$$\rightarrow $$^{3}P_{0}$,
  and $^{3}P_2$$\rightarrow $$^{3}P_{1}$ of  neutral atomic carbon (C)
  are available,  and find their  average excitation conditions  to be
  strongly subthermal.  This is  deduced from the CI(2-1)/(1-0) ratios
  ($\rm  R^{(ci)}_{21/10}$)  modeled by  the  exact  solutions of  the
  corresponding 3-level system, without  any special assumptions about
  the kinematic state  of the concomitant $\rm H_2$ gas  (and thus the
  corresponding line formation mechanism).  This non-LTE excitation of
  the CI lines  can induce the curious  clustering of (CI,LTE)-derived
  gas temperatures near $\sim $25\,K reported recently by Valentino et
  al.  (2020), which  is uncorellated to the  actual gas temperatures.
  The non-LTE  CI line excitation in  the ISM of galaxies  deprives us
  from a simple method for  estimating molecular gas temperatures, and
  adds uncertainty in CI-based molecular gas mass estimates especially
  when    the   J=2--1    line    is   used.     However   the    $\rm
  R^{(ci)}_{21/10}$=$\rm F(n,  T_{k})$ ratio is now  more valuable for
  joint CO/CI SLED and dust SED models of galaxies, and independent of
  the assumptions used in the  CO radiative transfer models (e.g.  the
  LVG approximation).   Finally we  speculate that the  combination of
  low  ratios  $\rm  R^{(ci)}_{21/10}$$\la  1$  {\it  and}  high  $\rm
  T_{dust}$ values found in  some extreme starbursts indicates massive
  low-density molecular wind and/or circumgalactic gas reservoirs.  If
  verified by imaging  observations this can be a  useful indicator of
  the presence of such reservoirs in~galaxies.

 \end{abstract}

\begin{keywords}
galaxies: ISM -- ISM: molecules -- (ISM:) cosmic rays -- ISM: atoms -- galaxies: starburst
\end{keywords}



\section{Introduction}

The  standard  Photon-Dominated Region  (PDR)  view  has the  CI  line
emission emanating  only from  a C-rich  transition layer  between the
outer  CII-rich  and  the  inner  CO-rich  regions  of  FUV-irradiated
molecular clouds, because  of a sharp recombination  of ionized Carbon
to neutral carbon and then to CO.  This was observed recently in small
regions near  the Be star $\rho$  Ophiuchi using ALMA but  in the same
image an extended  CI-rich molecular gas component is  hard to contain
in  the standard  PDR picture  (Yamagishi  et al.   2021).  The  first
observations challenging a PDR-only origin  of the CI line emission in
the ISM  were obtained by  Keene at al (1985,  1986) and Plume  et al.
(1994), showing the  CI $^{3}P_1$$\rightarrow $$^{3}P_{0}$ (henceforth
CI 1-0)  line emission  persisting well past  PDR fronts,  deep inside
CO-rich molecular clouds and even in a FUV-shielded dark cloud.  Since
then a widespread (CO,  $^{13}$CO)-C concomitance has been established
for the Orion A, B molecular clouds (Ikeda et al.  2002), the Galactic
Center  (Ojha  et  al  2001;  Tanaka et  al.   2011),  cold  dark  and
FUV-shielded clouds  in the Galaxy  (Oka et  al.  2001), and  a similar
such cloud  in M\,31  (Israel, Tilanus,  \& Baas  1998).  CI  1-0 line
emission  is  detected  in   very  FUV-shielded  regions  deep  inside
classical $''$dark$''$ clouds  such as TMC-1 (Tatematsu  et al.  1999)
and Bok  globules like Barnard  68. In the latter  the CI 1-0  line is
brighter  than  what PDR  models  that  fit  the CO,  $^{13}$CO  lines
predict, and Cosmic Rays (CRs) seem necessary to explain the molecular
gas thermal state (Pineda \& Bench~2007).

Nevertheless,  theoretical   prejudices  $''$conspired$''$   with  the
demanding nature of  the CI 1-0 line observations at  492\,GHz and the
lack of multi-beam receivers capable for  rapid CI line mapping in the
few  places  on Earth  where  such  observations are  possible  (Mauna
Kea/Hawaii, the  Atacama desert Plateau  in Chile, and  Antarctica) to
keep  intact the  notion  of CI  lines emanating  only  from a  narrow
CII/CI/CO transition layer  in the outer regions  of molecular clouds.
A  comprehensive  theoretical  and   observational  case  for  a  C/CO
concomitance  in  molecular  clouds,  with  CI  lines  an  alternative
molecular gas mass  tracer over large scales,  was made (Papadopoulos,
Thi, \&  Viti 2004) and boils  down to: a) cosmic  rays controling the
C/CO  abundance  throughout  the  FUV-shielded  volumes  of  molecular
clouds,  b)  a strong  turbulent  mixing  of any  initial  FUV-induced
CII/CI/CO species stratification, and  c) a non-equillibrium chemistry
operating  within  finite-age  molecular clouds  ($\rm  T_{age}$$\la$$
10^{7}$\,yrs: the chemical and  dynamical $''$recycle$''$ time for the
$\rm   H_2$$\rightarrow$HI    phase   transition).     In   vigorously
star-forming  galaxies  all three  factors  will  further enhance  the
[C/CO]  abundance throught  the  molecular  clouds.  Clumpy  molecular
cloud structures (e.g. Stutzki et al.   1998) can maintain some of the
observed  C and  CO  concomitance deeper  than  expected from  uniform
FUV-irradiated cloud  $''$surfaces$''$ without  fundamentally altering
the static  PDR carbon-species  stratification picture (Spaans  \& van
Dishoeck   1997),  but   are   hard-pressed  to   explain  the   tight
CI-($^{13}$CO,$^{12}$CO) {\it brightness  correlation} and the similar
CO,  C line  profiles observed  over a  wide range  of ISM  conditions
(Papadopoulos et al.  2004).

Recent theoretical work  on the role of CRs  (Bisbas, Papadopoulos, \&
Viti  2015;  Bisbas  et  al.    2017),  as  well  as  turbulent  cloud
simulations incorporating  the chemical  network controlling  the C/CO
abundance (Offner  et al. 2014; Glover  et al.  2015; Glover  \& Clark
2016; Clark et al. 2019) have largely borne this picture out.  However
it was  the large accumulation  of CI  line data in  the extragalactic
Universe, that  marked a true advance  of our views about  the regions
and conditions where these two lines emanate in molecular gas.

The Herschel  Space Observatory (HSO)  yielded large CI  line datasets
for Luminous  Infrared Galaxies  (LIRGs) in  the local  Universe (e.g.
Pereira-Santaella et al.  2013; Schirm et al.  2014; Kamenetzky et al.
2014; Papadopoulos  et al.   2014; Israel, Rosenberg  \& van  der Werf
2015; Lu  et al.  2017; Crocker  et al.  2019), while  earlier efforts
took  advantage of  large  redshifts  bringing the  CI  lines to  more
favorable  atmospheric windows,  to detect  starburst and  AGN in  the
distant  Universe  (Weiss et  al.   2003,  2007;  Walter et  al  2011;
Schumacher et  al.  2012;  Can\~ameras et al.   2018; Nesvadba  et al.
2019).  The onset  of ALMA operations at the   dry high plateau of
the Atacama desert generated spectacular amounts of CI line data both
locally (Cicone et al.  2018; Miyamoto et al.  2018; Jiao et al.
2017, 2019; Salak  et al.  2019; Michiyama et al.   2020; Saito et al.
2020; Izumi et  al.  2020; Miyamoto et al.  2021;  Dunne et al.  2021),
and  in the  distant  Universe (e.g.   Alaghband-Zadeh  et al.   2013;
Bothwell et  al.  2017;  Popping et  al.  2017;  Emonts et  al.  2018;
Andreani et al.  2018; Lu et al.  2018; Banerji et al.  2018; Lelli et
al.  2018; Valentino et al. 2018, 2020; Man et al. 2019; Harrington et
al. 2021; Boogaard et al. 2020).

The  aforementioned   observations  paint  a  consistent   picture  of
concomitant C and CO, with CI  1-0 a viable alternative $\rm H_2$ mass
tracer at the scales examined.   Nevertheless there are cases with the
CI (1-0)  line remaining undetected in  CO-detected regions (Michiyama
et al. 2020;  Miyamoto et al.  2021), and the  other way around (Dunne
et  al.   2021).   More  observational work  is  necessary  for  these
outliers to probe  the possibility of very different  {\it global} ISM
conditions, and  address possible sensitivity issues  in detecting the
fainter CI 1-0  emission expected in CO-rich but cold  and denser gas.
In the most  recent CI 1-0, 2-1 line dataset  analysed by Valentino et
al.  2020 for Main Sequence galaxies at z$\sim $1 the (CI,LTE)-derived
gas      temperatures      cluster       around      $\langle      \rm
T_{k}(CI,LTE)\rangle$$\sim $26\,K, are $\rm  \la T_{dust}$, and have a
dispersion of $\rm \sigma(T)$$\sim $8\,K.  If these are actual average
$\rm T_{kin}$ values for the molecular gas, they are hard to reconcile
with typical  (CO SLED)-derived $\rm  T_{kin}$ values for  SF galaxies
in the  local Universe  where $\rm T_{kin}$$\sim$(30--100)\,K  (e.g.
  Wall et  al.  1993; Aalto  et al.  1995; Papadopoulos et  al. 2010a;
  Kamenetzky et al. 2014).

In this  work the largest  extragalactic CI  2-1, 1-0 line  dataset to
date is used  to examine the average CI line  excitation conditions by
using: a) the exact analytical solutions of the CI 3-level system, and
b) the dust  temperatures obtained from dust  emission Spectral Energy
Distributions (SEDs) available for our sample. We find the CI lines to
be globally subthermally  excited in the ISM of  galaxies, discuss its
impact  on  molecular gas  temperature  and  mass estimates,  and  the
utility  of  a non-LTE  CI  2-1/1-0  ratio $\rm  R^{(ci)}_{21/10}=F(n,
T_{kin})$  in joint  models with  global CO  SLEDs, and  dust SEDs  of
galaxies.   Finally  we  highlight   the  possibility  of  using  $\rm
R^{(ci)}_{21/10}$ versus the $\rm T_{dust}$ extracted from global dust
emission SEDs  as an indicator  of massive molecular gas  winds and/or
circumgalactic gas reservoirs around starburst galaxies.

\section{The CI line and dust continuum data}

We canvassed the literature for all  available CI 2-1, 1-0 line data for
galaxies  that also  have submm/FIR  dust continuum  data adequate  to
provide an estimate of the luminosity-weighted $\rm T_{dust}$ from the
fitting  of a  modified black-body  (MBB).  This  yields 106  sources,
ranging from  nearby quiescently  star forming galaxies  (e.g. NGC891)
and  local (U)LIRGs  to  high redshift  lensed submillimeter  galaxies
(SMG)        and        QSO        with       a        range        of
\Lir$\sim   $$(10^9-5\times10^{13})\,\lsun$   (values  corrected   for
lensing).   Using  conversion  between  SFR  and  \Lir  of  $1.5\times
10^{-10}$, Kennicutt  \& Evans 2012;  Hao et  al. 2011; Murphy  et al.
2011), this  corresponds to SFR$\sim  $0.15--8000\msun\,$\rm yr^{-1}$,
i.e.  from  the very  quiescent to the  extreme starburst  SF activity
level. The sample  (presented in full in Dunne et  al.  {\em in prep})
is drawn  from the  following literature sources:  (Alaghband-Zadeh et
al. 2013; Aniano  et al. 2020; Chang  et al.  2020; Chu  et al.  2017;
Clark et al.  2018; Crocker et al. 2019; Jiao et al. 2017, 2019, 2021;
Garcia-Gonzalez et  al. 2016; Harrington  et al.  2021, Israel  et al.
2015; Kamenetzky et al.  2014; Lapham\&  Young 2019; Liu et al.  2015,
2021; Lu  et al.   2017, Nesvadba  et al.   2019; Papadopoulos  et al.
2010a;  Pereira-Santaella   et  al.  2013;  Rosenberg   et  al.  2015;
Schumacher  et  al.  2012;  Stacey  et  al.   2018; Valentino  et  al.
2018,2020; Walter et al. 2011).

In  this work  we use  a simple  but effective  argument based  on the
thermal decoupling of gas and dust  to set lower limits on the average
\tk\ of  the molecular  gas reservoirs  in the  sources of  our sample
using average  \td\ values (see Section~\ref{alpha}).   In this manner
we  avoid  the  well-known  difficulties of  determining  the  average
\tk\ of molecular gas which  requires well-sampled CO SLEDs, including
optically thin isotopologues, and tracers of high density gas in order
to  break  the \tk-n  degeneracies  inherent  in LVG  modelling  (e.g.
Papadopoulos et al.  2014). Needless to say that such difficulties are
massively  compounded  by the  size  of  our  sample  as well  as  the
non-uniformity of the  molecular, and atomic SLEDs  available for each
object.  Now  in order to  determine the  average \td\ values  the MBB
models  use  data  at  $\lambda_{\rm  rest}$>60\mic  ,  while  shorter
wavelength data,  if available, set  an upper  limit to the  warm dust
mass (at $\lambda_{\rm rest}$<60\mic the  dust SED diverges from a MBB
due to  transiently heated  small grains,  PAH and/or  hot AGN-powered
components).  We  use $\beta$=1.8 for  the fits, where  the literature
source provided  a fit using  a similar  method we use  their results,
otherwise      we     re-fitted      the     published      photometry
ourselves\footnote{Other sources  of  FIR   data:  Chu  et
  al. (2017), Clark et al. (2018).}.

We recognize that one-component dust continuum fits are overly simple,
and for global (i.e.  spatially  unresolved) SEDs tend to be dominated
by the  warmer dust compoment(s).   They are however adequate  for our
purpose of using the deduced luminosity-weighted $\rm T_{dust}$ values
to  normalize the  $\rm T_{kin}$(CI,\,  LTE) values  extracted from  a
similarly one-component modeled $\rm  R^{(ci)} _{21/10}$ ratio. This
  is because  even for modestly  FUV-irradiated/CR-innudated molecular
  clouds in SF galaxies, gas with n$\la $$10^4$\,$\rm cm^{-3}$ will be
  CI-rich and warm  (Papadopoulos, Thi, \& Viti  2004; Papadopoulos et
  al.  2011) while containing also the bulk of the gas (and thus dust)
  mass per molecular cloud\footnote{Even  for the highly turbulent ISM
    of merger/starburst systems observations  show at most $\sim $25\%
    of molecular gas  mass to reside at  densities $>$$10^4\,\cc$ (Gao
    \& Solomon 2004).}.   The collective emission of  such clouds will
  thus dominate  the global  CO, CI  SLEDs and  dust emission  SEDs in
  galaxies (a possible large scale deviation from this simple picture
  is discussed in Section~\ref{windCG}).

\subsection{The cases of high dust optical depth.}
\label{ODS}

Some compact starbursts  and/or deeply dust-enshrouded AGN  in our sample
may have dust  SEDs that are optically thick well  into the FIR/sub-mm
(e.g. Arp220, GN20: Papadopoulos 2010b;  Scoville et al. 2015; Cortzen
et al. 2020). This has two main effects:

\begin{enumerate}
\item{The  isothermal  luminosity-weighted  dust  temperature,  fitted
  under the assumption of  optically thin emission under-estimates the
  true $\rm  T_{dust}$ in  an optically  thick dust  SED as  the submm
  $''$excess$''$ due to high dust optical depths will be attributed to
  colder dust.   Thus the  true \td($\tau$) will  be higher  than from
  that obtained under an optically thin MBB.}

\item{The observed ratio of \CI 2-1/1-0 is suppressed by the differing dust
  optical depths between the two frequencies and must be corrected in
  order to recover the intrinsic ratio, and hence the intrinsic
  \tx.}

\end{enumerate}

This correction is discussed in Papadopoulos (2010b) and can be expressed as:

  \begin{equation}
\rm R^{(ci)}_{21/10}(int)= R^{(ci)}_{21/10}(obs)\times \exp\left[\tau_d(\nu_{21})-\tau_d(\nu_{10})\right]
  \end{equation}

\noindent
where $\tau_d(\nu)$ is the dust optical depth at the frequencies
$\nu_{21}$ (809~GHz) and $\nu_{10}$ (492~GHz) of the \CI lines, and

\begin{equation}
\rm R^{(ci)}_{21/10}= \frac{\rm \int_{\Delta V} S_{CI}(2-1)dV}{\rm \int _{\Delta V} S_{CI}(1-0)dV},
\end{equation}

\noindent
is the  CI (2-1)/(1-0) line  ratio ($\rm \int_{\Delta V}  S_{CI}dV$ in
Jy\,km\,s$^{-1}$: the  velocity-integrated line flux  density), with
(int)   and  (obs)   designating  the   intrinsic  (i.e.    $\rm  \tau
_{d}$-corrected)  and observed  values respectively.   The corrections
obtained  for  the optically  thick  sources  are $\sim$1.2--2.8  (the
highest  value in  Arp220), and  shift  the CI  line ratios  of a  few
optically  thick  sources  (GN20,  IRASF\,10214,  AMS\,12)  to  higher
corresponding densities and almost into the LTE excitation domain (see
discussion in ~\ref{nonLTEratio})

The dust SED fits are {\em highly degenerate} between solutions with a
higher  dust  optical  depth,  and   those  with  a  larger  range  of
temperatures  contributing to  the  SED.  We  therefore  only use  the
optically  thick parameters  for  6 local  (U)LIRGs  where other  
evidence  supports  very  compact  bright  regions producing  the  IR
emission (e.g.   interferometric imaging showing very  high mm surface
brightness,  evidence for  X-ray  absorbed AGN,   severe deficit  of
C[{\sc ii}] relative  to \Lir, or high brightnesses  of emission lines
(e.g. HCN-{\em vib}) that are only  excited in the presence of extreme
IR radiation fields (Gonz\'alez-Alfonso \& Sakamoto 2019, Boettcher et
al. 2020;  Falstad et al. 2021).   Not all these sources  have extreme
\Lir;  NGC~4418 and  Zw\,049.057  are examples  of moderate  \Lir$\sim$
$2\times10^{11}\,\lsun$  sources  with  very  dense  and  compact  dust
regions  ($\rm{N_H\ga10^{25}}\,\cc$)).\footnote{The  observed \tx  of
  these  sources  are very  low  13--18~K, further supporting  the
  necessecity of optical depth corrections.}   Additionally, we use
optically thick corrections to \CI\  line ratios for 16 high-z SMG/QSO
which were fitted with $\td(\tau)$$>$50\,K,and had enough data points to
exclude a simple isothermal optically thin SED. In total we apply such
corrections to 22  sources, and show the results with  and without the
optical depth corrections (see~\ref{nonLTEratio}).

\subsection{The non-LTE excitation of the CI 1-0, 2-1 lines}
\label{subtherm}

\begin{figure}
  \includegraphics[width=0.9\columnwidth]{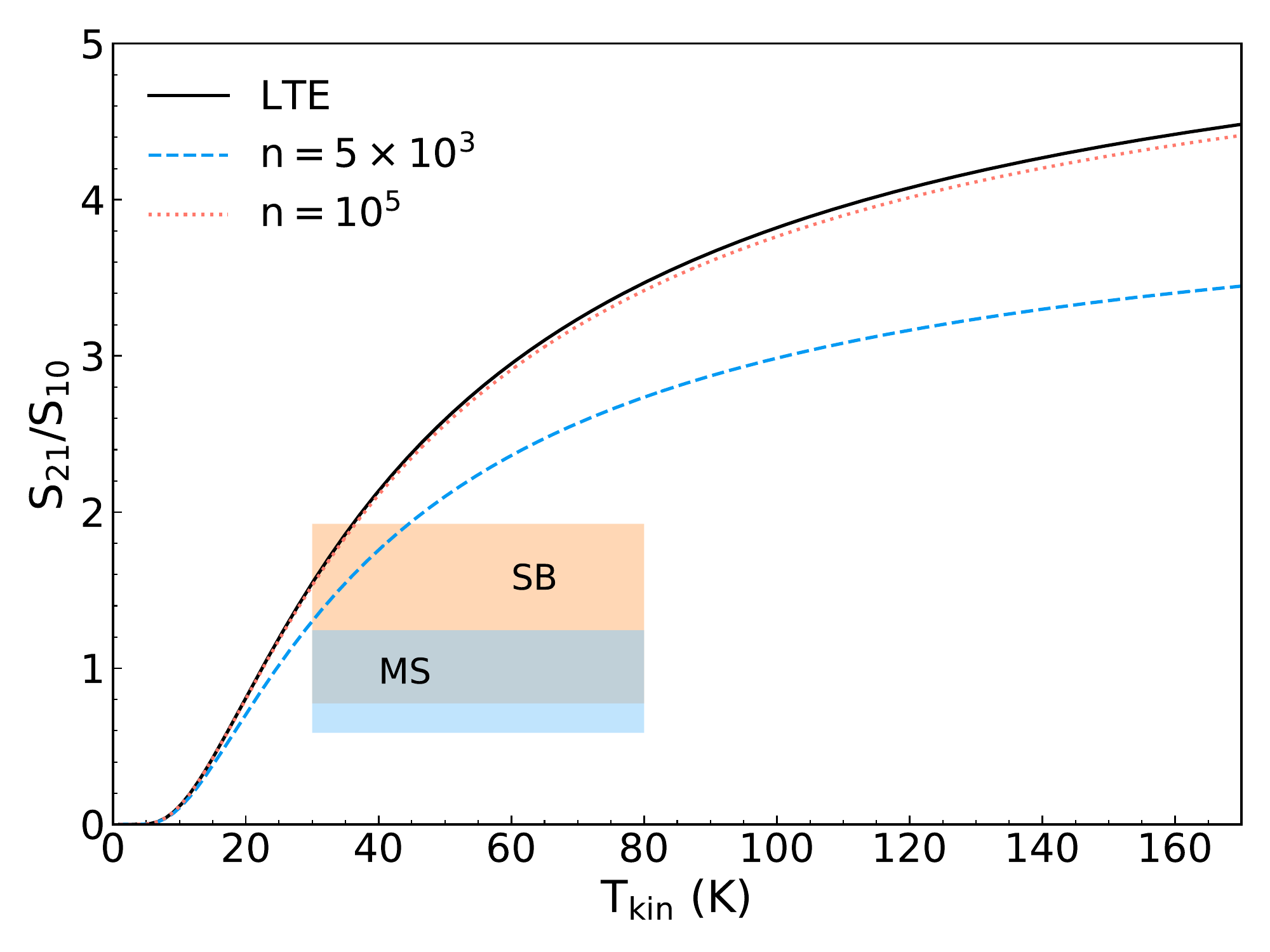} 
\caption{\label{RcitkF} The $\rm R^{(ci)}_{21/10}$=$S_{21}/S_{10}$=$\rm F(T_{kin}, n)$ line ratio
  ($\rm S_{21}$, $\rm S_{10}$ in Jy\, km\, s$^{-1}$)
  versus \tk, computed using the analytical  non-LTE expressions for a 3-level system, in the optically thin
  regime.  We show 3 cases: full LTE (black line), n=$10^5 \cc$
  (red dotted line), and n=5$\times $$10^3\cc$ (blue dashed line). The
  critical density of the \CI(2-1) line $\rm n_{21,crit}$$\sim$
  $10^{3}\,\cc$.  {\it The shaded regions:} (vertical)  the $1\sigma$ range
  of the $\rm R^{(ci)}_{21/10}$ values in our sample (salmon: star-bursts, blue:
  main-sequence (MS) galaxies), (horizontal) the typical \tk\ range deduced from CO SLED
  modelling in literature studies.}
\end{figure}

  For the CI lines it is: $\rm T_{1}$=$\rm E_{1}/k_B$=23.6\,K and $\rm
  E_{2}/k_B$=62.4\,K, while the $\rm T_{kin}$ values from SLED fits of
  extragalactic   CO    lines   are   typically    $\rm   T_{kin}(CO\,
  SLED)$$\sim$(30-100)\,K  (i.e.   nearly  $''$bracketed$''$  by  $\rm
  E_1/k_B$ and $\rm E_2/k_B$).  A  purely-LTE CI line excitation, with
  $\rm  T_{kin}$(CI, LTE)$\sim  $$\rm  T_{kin}$(CO  SLED), should  then
  yield  a   $\rm  R^{(ci)}_{21/10}$  ratio  that   is  strongly  $\rm
  T_{kin}$-variable   throughout   this   $\rm   T_{k}$   range.    In
  Figure~\ref{RcitkF}  we  plot  the  analytical  expression  of  $\rm
  R^{(ci)} _{21/10}$=$\rm  F(n, T_{kin})$  (using formulas  taken from
  Papadopoulos, Thi \& Viti 2004, Appendix)  where it can be seen that
  for    $\rm   T_{kin}(LTE,    CI)$$\sim   $30--100\,K)    the   $\rm
  R^{(ci)}_{21/10}(LTE)$  ratio  indeed  varies  significantly  ($\sim
  $1--3.8),  but   also  that   most  of   the  {\it   observed}  $\rm
  R^{(ci)}_{21/10}$  values,  lie  well below  the  corresponding  LTE
  values.

An  educational aspect  of Figure  1  is the  demonstration that  $\rm
R^{(ci)}_{21/10}$  remains  below  its   purely-LTE  values  even  for
densities as  high as  n=$5\times 10^3$\,cm$^{-3}$($\sim  5\times$ the
critical  density  of   CI  2-1).   This  means   that  spectral  line
thermalization is  a very  gradual function  of $\rm  n/n_{crit}$, and
that the often encountered assertion  that a spectral line reaches LTE
conditions  when   gas  densities   reach  n$\sim$$\rm   n_{crit}$  is
incorrect.

Finally  from Figure  1  we  can see  also  that  the gas  temperatures
corresponding to the measured $\rm R^{(ci)}_{21/10}$ line ratios under the
LTE assumption, namely:

\begin{eqnarray}
  \rm T_{kin}(CI, LTE)&=&
  \rm \left(E_{21}/k_B\right) \left[ln \left(\frac{g_2 A_{21}}{g_1 A_{10}} \frac{1}{R^{(ci)}_{21/10}} \right)\right]^{-1}\\ 
   \rm                  &=& \rm \frac{38.8\,K}{ln\left(5.63/R^{(ci)}_{21/10}\right) \nonumber}
\end{eqnarray}

\noindent
(for  $\rm  E_{21}/k_B$=38.8\,K, $\rm  A_{21}$=$2.68\times  10^{-7}$\,
s$^{-1}$, $\rm  A_{10}$=$7.93\times 10^{-8}$\, s$^{-1}$, Zmuidzinas
et al.  1988) will all gather in a  relatively narrow range  of $\rm
T_{kin}(CI, LTE)$$\sim $(20-30)\,K (if  one $''$projects$''$ the lower
and the upper line marking the 1-$\sigma$ width onto the black LTE curve, and then
$''$reads$''$ off the corresponding temperatures of the thus projected points).
This is another manifestation of non-LTE conditions which we discuss further
in Section~\ref{DIS}.

We  must  mention  that  piecemal observational  evidence  for  global
subthermal excitation  of CI lines  has been available for  some time.
The  earliest  examples of  $\rm  T_{kin}(CI,  LTE)$ ($\rm  T_{kin}  $
deduced from the CI lines under  the LTE assumption, Equation 3) being
lower than $\rm T_{kin}$(CO SLED) and $\rm T_{dust}$, were reported by
Kamenetzky et al.   (2014) and Schumacher et al.   (2012).  The former
found  $\rm  \langle  T_{kin}(CI,\,  LTE)\rangle$=$\rm  (29\pm  6)\,K$
$\la  $$ \rm  T_{dust}$, and  uncorrelated to  the $\rm  T_{kin}$(CO\,
SLED) values, while the later deduced $\rm T_{kin}$(CI,\, LTE)$<$($\rm
T_{dust}$,  $\rm  T_{kin}$(CO  SLED))  for  a  distant  quasar.   Both
interpreted  these temperature  differences  as due  to different  ISM
regions sampled by the CO, CI lines and the dust continuum rather than
large scale sub-thermal  CI line excitation.  Only in  the most recent
study of Lensed Planck-selected  (LP) galaxies by Harrington et al.
(2021) the  possibility of  global subthermal  CI line  excitation was
recognized,  although  their  argument   about  enhanced  gas  kinetic
temperatures  with  respect  to   carbon  excitation  temperatures  is~unclear.


\subsection{The CI line excitation, the detailed picture}

\begin{figure}
\includegraphics[width=0.5\textwidth, trim=0.4cm 0.4cm 0cm 0cm, clip=true]{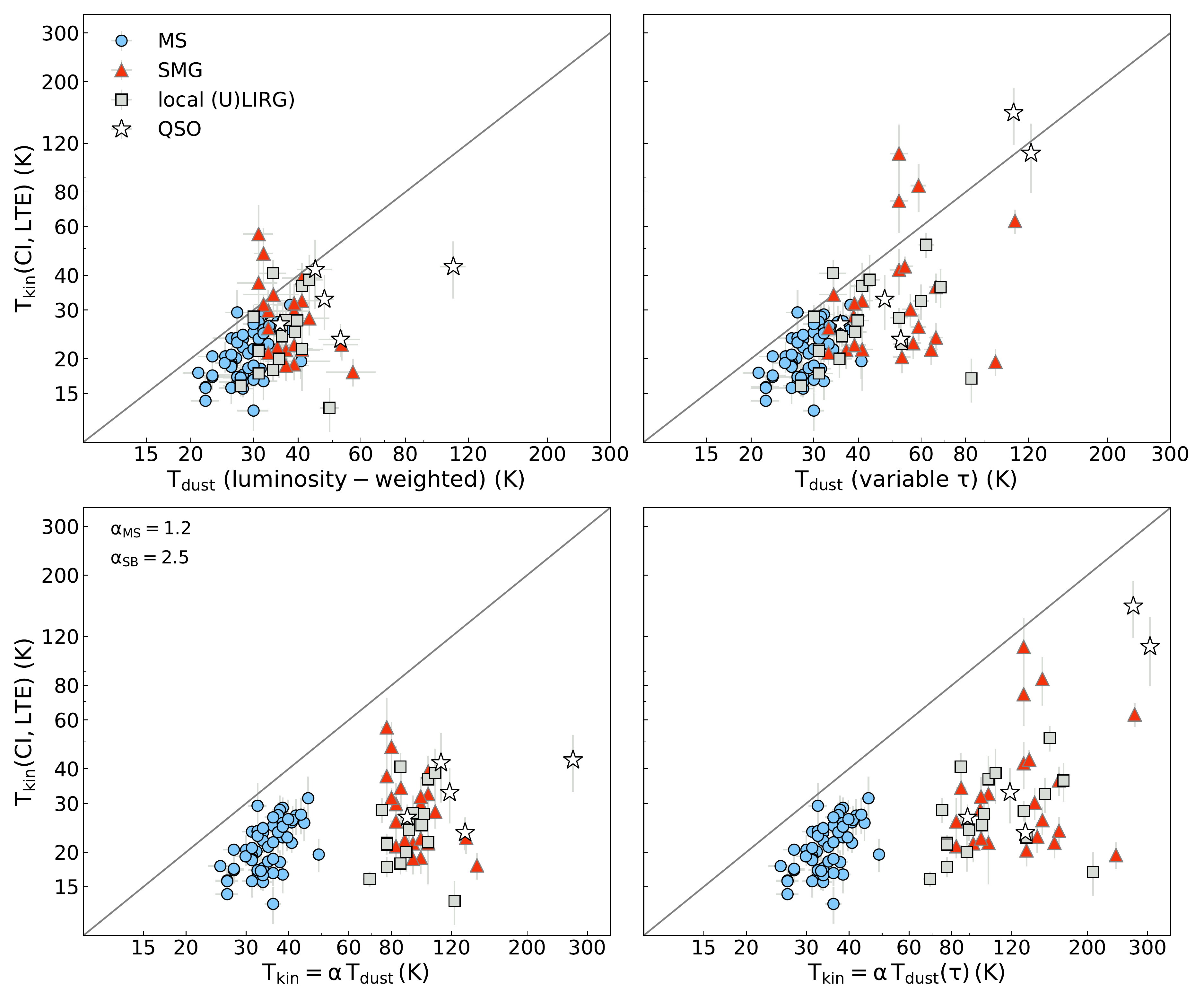}
  \caption{\label{txtdF} {\bf Top left:} \tx versus \td. The {\em
    minimum} values of $\tx$=$\td$ ($\alpha$=1) expected for concomitant gas and dust
  mass reservoirs (see Section~\ref{alpha}) are marked by a solid
  line. The luminosity-weighted dust temperature was obtained from the
  dust SEDs of our sample using optically thin MBB fits at
  $\lambda_r$>60\mic with $\beta$=1.8.  {\bf Top right:} As top left
  but correcting for sources which are plausibly optically thick in the
  FIR-submm part of the dust SED, by using dust SED fits that include
  a characteristic optical depth as a free parameter, and correcting
  $\rm R^{(ci)}_{21/10}$ (and hence \tx) accordingly
  (Section~\ref{ODS}) {\bf Lower left:} \tx versus the \tk\ expected for
  the computed \td, with a solid line marking \tx=\tk\ (which would be
  satisfied for the LTE regime).  We adopted $\rm \alpha_{MS}$=1.2 for
  typical SF galaxies on the `Main Sequence' (MS), and $\rm
  \alpha_{SB}$=2.5 for `extreme starbursts' (ULIRGs/SMG/QSO) (see
  Section~\ref{alpha}).  {\bf Lower right:} As in lower left, left but
  correcting for optically thick dust SEDs.}
\end{figure}

\begin{figure}
\includegraphics[width=0.48\textwidth,trim=0cm 0cm 0cm 0cm, clip=true]{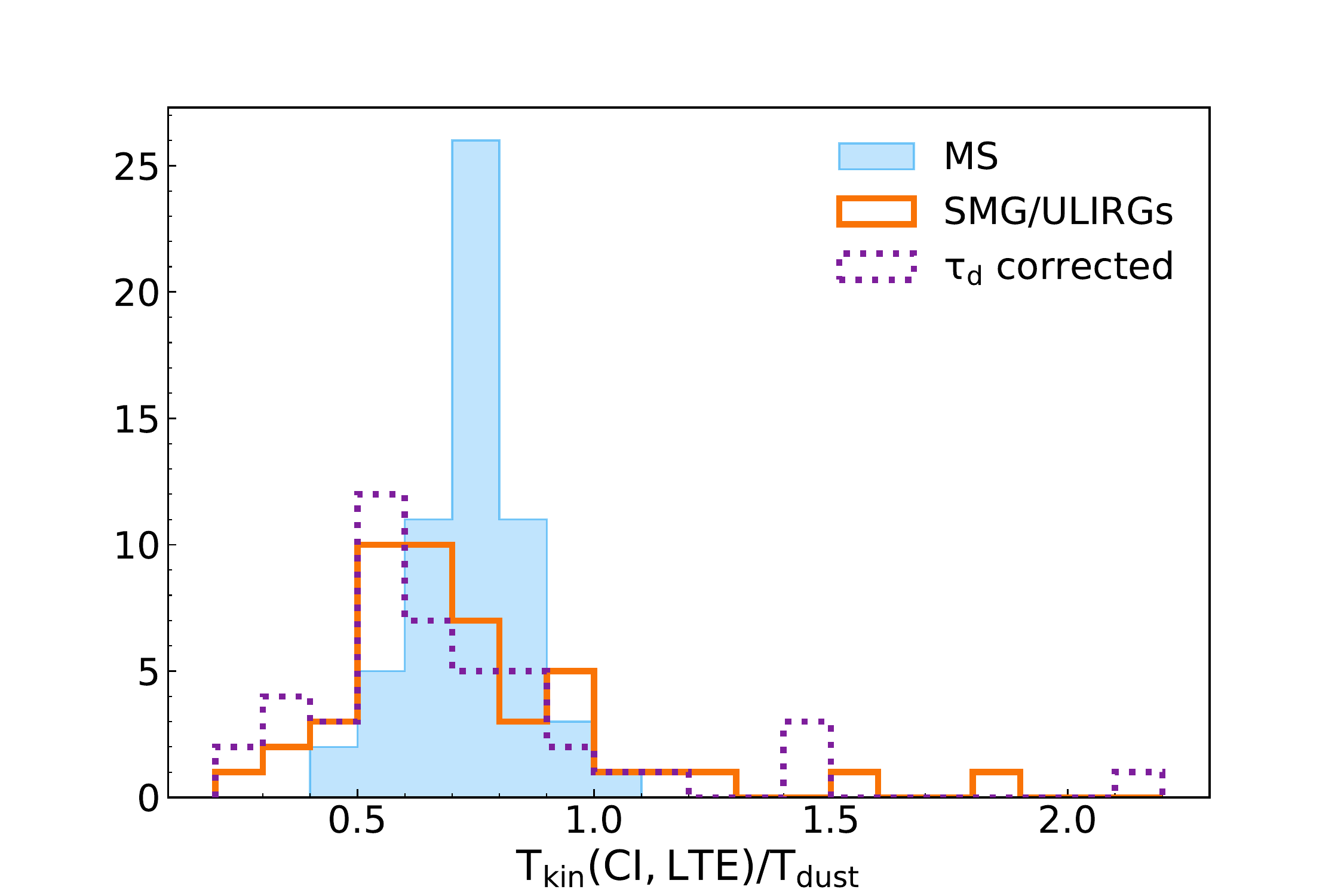} 
\caption{\label{txtdhistF} The distribution of the $\tx/ \td$ ratio,
  colour coded by the SF mode (normal=blue, extreme=orange). The
  purple dotted line represents the distribution for the extreme SF
  sources (SMG/(U)LIRGS) where corrections have been made to account
  for high optical depth in the dust SED of 22 sources (see
  Section~\ref{ODS}). There is no significant difference in the
  distribution of \tx/\td\ whether optically thick fits are considered
  or not. Overall the SMG/ULIRG sources have a wider spread of ratios,
  but both galaxy classes are significantly below the minimum value
  $\tk$=$\td$, with a mean $\langle \tx/\td \rangle=0.73\pm0.17$.}
\end{figure}

In Figures~\ref{txtdF}, ~\ref{txtdhistF} we show the $\rm T_{kin}$(CI,
LTE) versus  $\rm T_{dust}$,  and $\rm T_{kin}$=$\rm  \alpha T_{dust}$
($\alpha $$\geq  $1), the  gas temperature expected  for a  given $\rm
T_{dust}$  and  an  ISM  with   primary  gas  heating  mechanisms  the
photoelectric  effect induced  on dust  by  FUV radiation  (from O,  B
stars), cosmic rays (CRs),  and/or shocks (see section~\ref{alpha} for
details and the adopted values for parameter~$\alpha$).

 These figures  make it obvious  that the  {\it CI lines  are globally
   subthermally   excited  in   the  ISM   of  galaxies}   since  $\rm
 T_{kin}$(CI, LTE) is lower than  $\rm T_{dust}$, and lower still than
 the expected  $\rm T_{kin}$=$\rm  \alpha T_{dust}$ in  all but  a few
 objects in our sample.  For  starbursts, where higher $\alpha$ values
 are  expected  for  their   intensely  CR-irradiated  ISM,  the  $\rm
 T_{kin}$(CI, LTE)<$\rm T_{kin}$=($\rm \alpha T_{dust}$) inequality is
 even more  pronounced, indicating very subthermal  CI line excitation
 in their H$_2$ gas reservoirs (see also discussion in~\ref{windCG}).

 In retrospect the global subthermal CI  line excitation in the ISM of
 galaxies  should not  be  so surprising  since most  of  the mass  in
 molecular  clouds,  even  very  turbulent  ones,  lies  at  densities
 n$<$$10^4$\,cm$^{-3}$,  where the  CI(2-1) line  remains subthermally
 excited (see section~\ref{subtherm}).  Furthermore, for CR-irradiated
 clouds,  the [CI/$\rm  H_2$] abundance  can be  $\sim $(5-10)$\times$
 higher  in lower-density  (n$\leq$$\rm (few)\times  10^3$\,cm$^{-3}$)
 than  higher density  ($\geq$$10^4$\,cm$^{-3}$) gas  (Papadopoulos et
 al. 2011), making the lower  density (and typically the most massive)
 $\rm H_2$  gas phase the  most C-rich one in Giant  Molecular Clouds
 (GMCs)\footnote{Moreover  lower  density  gas is  also  warmer  gas
   (since dominant  cooling has $\Lambda $$\propto$$\rm  n^2$) further
   enhancing the contribution of the lower  density gas to the CI line
   luminosities.}.

\subsubsection{A note on the  $\rm T_{kin}\ga T_{dust}$ gas-dust thermal decoupling, and the
choice of $\alpha$ }
\label{alpha}

A key assumption  used in this work is that  $\rm T_{kin}$=$\rm \alpha
T_{dust}$, with  $\alpha \geq 1$ holds  for most of the  molecular gas
mass in  galaxies.  This is  both theoretically expected  and actually
observed.  Early PDR models  yielded this inequality in FUV-irradiated
clouds  (e.g.   Hollenbach  \&  Tielens 1999,  Figure  16)  with  $\rm
T_{kin}$$\sim$$ \rm T_{dust}$ (i.e gas-dust thermal coupling) occuring
only at $\rm A_{v}$$\ga $6--10 in small dense regions deep inside $\rm
H_2$ clouds,  where both gas and  dust are cold. Deeper in such
  regions $\rm T_{kin}$ can become lower  than $\rm T_{dust}$ by a few
  Kelvin (e.g.  Papadopoulos et al.   2011; see also Hollenbach et al.
  1991  for   earlier  seminal  work)   which,  for  the   purpose  of
  extragalactic line and dust  emission models, are essentially equal.
  Furthermore such  $\rm T_{dust}$$\ga $$\rm T_{kin}$  regions will be
  inconspicious in  the emergent   (H$_2$ cloud)-scale  (and thus
  galaxy-scale)  line   and  dust  continuum  luminosities,   even  in
  SF-quiescent spirals like the Milky Way with much of their ISM already
  cold  ($\sim  $20\,K).  This  is  simply  because such  regions  are
  typically colder still  ($\sim $8-10\,K), and contain  only few\% of
  an   H$_2$  cloud's~ mass\footnote{when   imaged   to  be   studied
    individually these CR-dominated regions are very important as
    the sites where  the initial conditions of star  formation are set
    (Bergin \& Tafalla 2007).}.

Beyond   the  FUV-induced   photoelectric  heating   maintaining  $\rm
T_{kin}$$\ga  $$\rm  T_{dust}$  throughout  most of  $\rm  H_2$  cloud
volumes, it is the {\it volumetric} cosmic ray (CR)-heating of the gas
(but  not  the  dust)  that decisively  sets  a  $\rm  T_{kin}$$>$$\rm
T_{dust}$  inequality throughout  H$_2$  clouds  (Papadopoulos et  al.
2011).  Even for modest  CR energy density enhancements, corresponding
to $\sim  $10$\times$ higher SFR densities  than in the Milky  Way, CR
heating can maintain $\rm T_{kin}$$>$$\rm T_{dust}$ for densities even up
to  $\sim  $$10^5$\,cm$^{-3}$ (and  thus  for  the  bulk of  an  H$_2$
cloud's~mass).

Finally we note that turbulence,  like CRs (and unlike FUV radiation),
will {\it volumetrically}  heat the gas but not the  dust. Thus it can
only  strengthen  the  original  FUV/CR-induced  $\rm  T_{kin}$$>$$\rm
T_{dust}$  inequality  throughout   molecular  clouds.   Observational
evidence for this,  using multi-species, multi-J SLEDs  and submm dust
continuum maps,  is available for  the turbulent molecular gas  in the
Galactic Center  (Papadopoulos et  al.  2014 and  references therein).
In the  highly turbulent  ISM of  extreme starbursts,  typically
  driven by mergers, the shock-dominated  turbulent gas heating can be
  significant and perhaps even dominant.

For  ordinary  SF systems  up  to  the  so-called Main  Sequence  (MS)
galaxies we adopt $\alpha$=1.2, the minimum value expected for gas and
dust temperatures deep inside  FUV/CR-irradiated clouds (see Figure 2,
Papadopoulos et  al.  2011), which  can actually be higher  ($\sim $2)
for the  CR energy densities  expected for the  upper SFR range  of MS
galaxies.   For starbursts  (SFR>100\, $\rm  M_{\odot}\, yr^{-1}$)  we
adopt $\alpha $=2.5,  a value deduced recently for  such galaxies from
joint  multi-J CO/CI  SLED  and dust  SED fits  by  Harrington et  al.
(2021), and readily supported by detailed thermo-chemical calculations
for the  strongly FUV/CR-irradiated molecular clouds  expected in such
intense starburst environments (Papadopoulos  et al.  2011). Here
  we must  note that the  choice of  the two representative  $\alpha $
  values is  driven merely by  the need to demonstrate  the subthermal
  excitation of  the CI lines  in a simple and  straightforward manner
  (while   bypassing  the   complexities   and   degeneracies  of   CO
  SLED-derived $\rm T_{kin}$ values).   In practice even within galaxy
  regions, let  alone entire galaxies  with different SFR  levels, the
  $\alpha  $  factor  will  change  as  a  function  of  $\rm  \langle
  n(H_2)\rangle $,  $\rm \langle U_{CR} \rangle$  (CR energy density),
  $\rm G_{\circ}$ (FUV field strength)  and $\rm \langle M_s \rangle $
  (average Mach number).  If well-sampled molecular (e.g. CO, HCN, and
  their  isotopologues),  and atomic  (CI)  SLEDs  and dust  SEDs  are
  available, the  average $\alpha $  values can actually  be extracted
  from  the  models  (see  Harrington   et  al.   2021  for  a  recent
  example). In  turn, such $\alpha $  values can be used  to determine
  the dominant  power source of the  ISM in the galaxy  observed (i.e.
  the Y factor)  with large $\alpha $ values  indicative of volumetric
  (CR and/or turbulence) rather that  FUV heating (see Papadopoulos et
  al.  2014 and references therein).   We will briefly return to these
  issues at Section~\ref{DIS}.






\subsection{The non-LTE $\rm R^{(ci)}_{21/10}$=$\rm F(n, T_{kin})$ ratio}
\label{nonLTEratio}

The subthermal  excitation of the CI  lines must be partly  the reason
why they are often faint in galaxies, even as the CI 1-0 line has been
successfully used as a global molecular gas tracer yielding gas masses
in  accordance with  those  deduced  from the  dust  emission (and  an
assumed dust/gas ratio) and CO lines (e.g.  Lee et al.  2021; Dunne et
al.  2021).  Under non-LTE conditions  the CI 2-1 line luminosity will
of course  be much more  sensitive to  the average $\rm  (n, T_{kin})$
values in  the ISM of  galaxies than the  CI 1-0 luminosity,  which is
what may  be allowing the  latter, even when  faint, to remain  a good
molecular gas tracer (more details on this in Section~\ref{DIS}).

The non-LTE CI line excitation in the ISM of galaxies readily deprives
the  $\rm  R^{(ci)}_{21/10}$ ratio  of  its  use  as a  molecular  gas
$''$thermometer$''$   since  now   $\rm  R^{(ci)}_{21/10}$=$\rm   F(n,
T_{kin})$. The latter however makes the  CI lines much more useful for
studies of  molecular gas  conditions.  Moreover unlike  the optically
thick CO rotational lines  with their infinite-term partition function
which necessitate:  a) radiative  transfer models  with an  upper $\rm
J_{up}$  cuttof, b)  assumptions  about the  line formation  mechanism
(LVG) and the corresponding expression for the line optical depth, and
c) solving iteratively for  level populations controlled by collisions
{\it and} the  local CO line radiation field itself,  {\it the 3-level
  system of the two fine structure  lines of the neutral atomic carbon
  can be  solved analytically.} Thus  the full non-LTE  expression for
$\rm R^{(ci)}  _{21/10}$=$\rm F(n,  T_{kin})$ can be  obtained without
assumptions  about  gas  velocity  fields and  their  impact  on  line
formation and optical~depth.

The  latter is possible only  for optically thin CI  lines, which
  early observations  of dense  molecular clouds showed  it to  be the
  case (Zmuidzinas et al.  1988), as well as in the bulk of the ISM in
  galaxies (Weiss et al 2003 and references therein; Harrington et al.
  2021).  Low line  optical depths is among the traits  that allow the
  CI  1-0  line to  trace  molecular  gas  mass (CI-based  H$_2$  mass
  estimates are always made under the optically thin line assumption),
  on par  with other optically thin  tracers such dust and  $ ^{13}$CO
  line emission.  Regarding the latter, the good spatial and intensity
  correlation between  CI 1-0 and  $ ^{13}$CO J=2--1 line  emission in
  molecular clouds (an important early  landmark in the case against a
  PDR-only origin  of CI line emission,  Keene et al.  1997)  would be
  difficult  to  achieve  unless  both  lines  were  mostly  optically
  thin\footnote{Should any cases of optically thick  CI lines
    over large  molecular gas  mass reservoirs  be found  in galaxies,
    detailed line radiative transfer in macroturbulent media must then
    be employed.  CI line emission models, while no longer analytical,
    would  nevertheless  still be  simpler  than those  for  the
    infinite J-ladder of CO lines.}.

\begin{figure*}
 \includegraphics[width=0.48\textwidth]{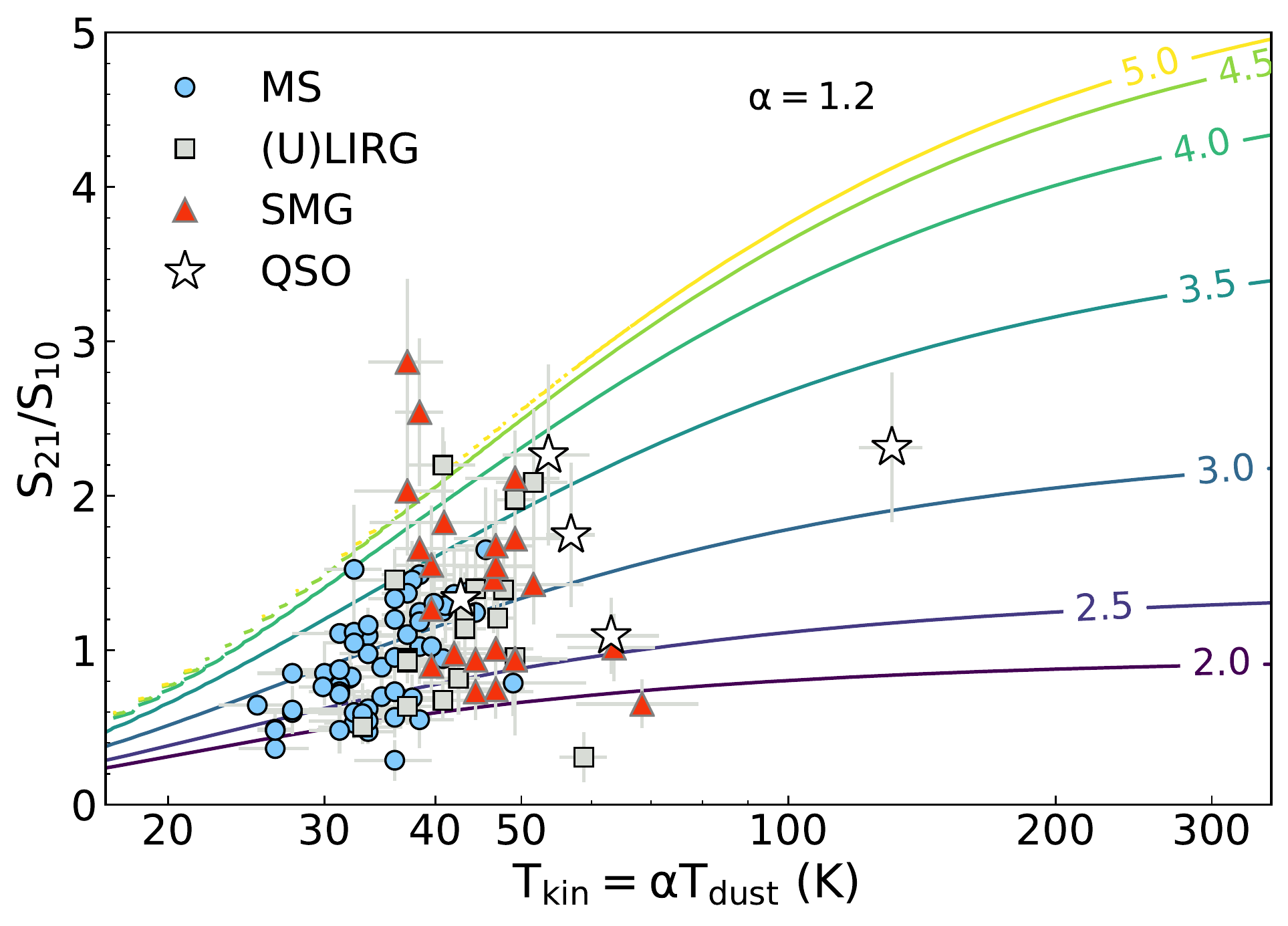}
\includegraphics[width=0.48\textwidth]{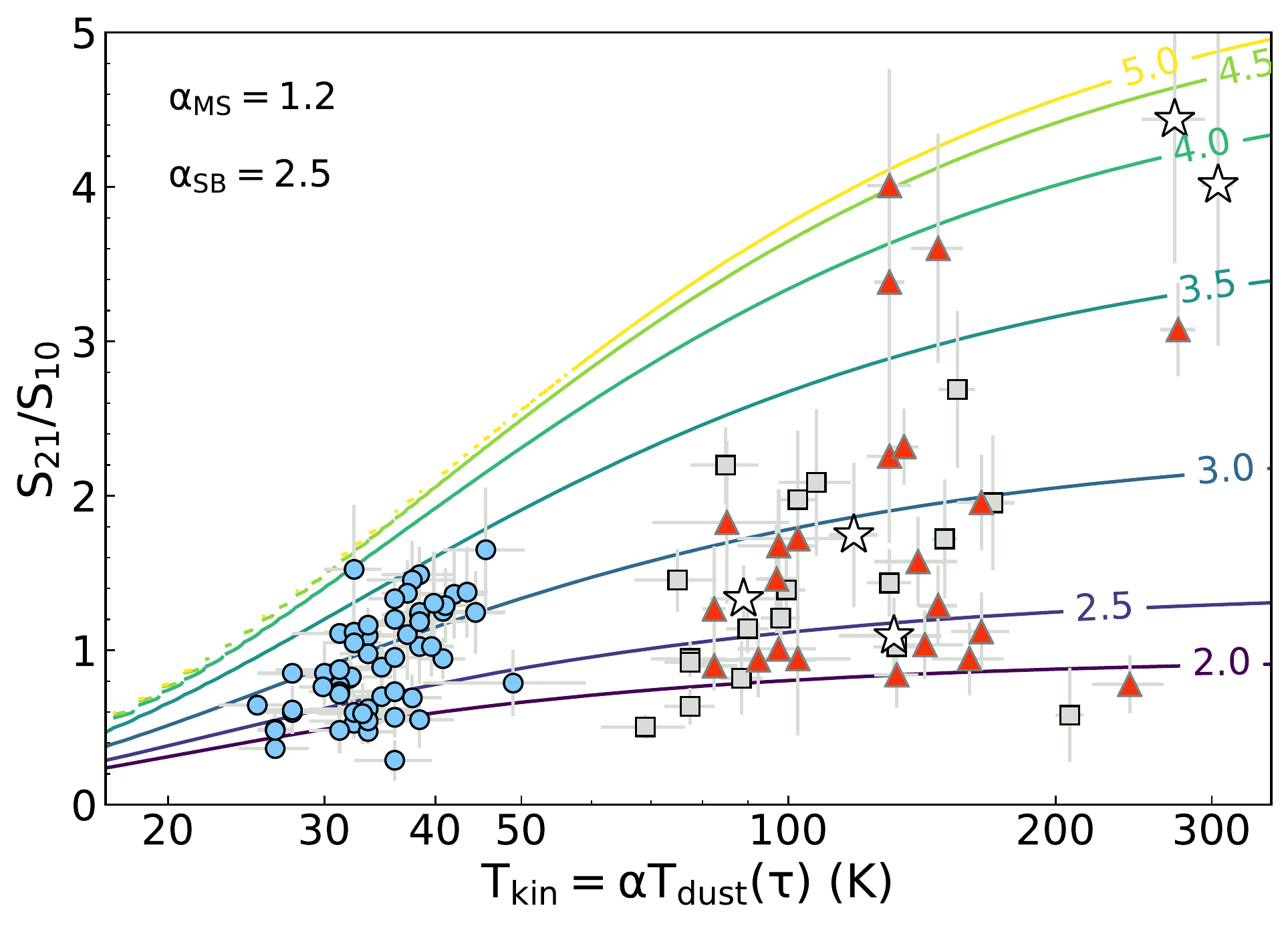} 
\caption{\label{R21tdF} The $\rm R^{(ci)}_{21/10}$ ratio versus
  \tk\ for a range of gas densities (shown as iso-density lines
  labelled by the corresponding log($n$) values).  The data from our
  sample are over-plotted using the same symbols as Figure~\ref{txtdF}:
  blue circles (Main Sequence galaxies), red triangles (high-z SMG),
  stars (QSO) and grey squares (local (U)LIRGS). {\bf Left panel:} We
  used the luminosity-weighted \td\ to derive \tk\ using constant
  gas-dust thermal decoupling factor $\alpha$=1.2
  (Section~\ref{alpha}). {\bf Right panel:} Using the optically thick dust
  SED fit results where appropriate to derive \td\ (and hence \tk=$\alpha$\td) and to
  correct the measured $\rm R^{(ci)} _{21/10}$ for the effects of the dust optical depth, see
  Section~\ref{ODS}). In this panel we use a higher value of
  $\alpha$=2.5 for the extreme star-bursts as deduced by Harrington et
  al. (2021) for such galaxies.}
\end{figure*}

The non-LTE expressions for the CI 3-level populations as functions of
density,  temperature  and  background  radiation field  are  used  to
compute the  $\rm R^{(ci)}_{21/10}$=$\rm F(n, T_{kin})$  function (see
Papadopoulos,  Thi  \& Viti  2004,  Appendix),  and  then plot  it  in
Figure~\ref{R21tdF}.    There    we   overlay   the    measured   $\rm
R^{(ci)}_{21/10}$ values,  to which  we assign the  corresponding $\rm
T_{kin}$=$\rm  \alpha T_{dust}$  for  the  concomitant molecular  gas,
using the  $\rm T_{dust}$  obtained for the  particular object,  and a
gas-dust thermal  decoupling factor $\alpha$  (see~\ref{alpha}).  Here
it is worth pointing out that  the value $ \alpha$=1.2 adopted for the
entire sample  (Figure~\ref{R21tdF}, left  panel) is  almost certainly
too low  for the ISM of  starburst galaxies whose actual  average $\rm
T_k$ will be higher.  In that regard the characteristic density of the
iso-density    curve    crossing    through   a    particular    [$\rm
  (R^{(ci)}_{21/10})_{meas}$,  $\rm   T_{kin}$]  point  in   our  $\rm
R^{(ci)}_{21/10}=F(n,  T_{kin})$ diagram  (Figure~\ref{R21tdF}) is  an
upper limit to the average density  of the molecular gas. In the right
panel where a higher value $\alpha $=2.5 was adopted for the starburst
galaxies  in  the  sample,  the   corresponding  points  move  to  the
lower-n/higher-$\rm T_k$ non-LTE part of the diagram.


 Figure~\ref{R21tdF}    shows    again     that,    for    n$\sim$$\rm
 n_{ik,crit}$=$\rm  A_{ik}/(\sum_{k}  C_{ik})$   (sum  of  collisional
 coefficients $''$running$''$ for both k$<$i {\it and} k$>$i values of
 k)  the  corresponding  i$\rightarrow$k transition  does  not  become
 thermalized  (where $\rm  T_{ex}(i,k)$$\sim$$\rm  T_k$).  Indeed  for
 n=$\rm  n_{2-1,  crit}$$\sim$$\rm 10^3\,  cm^{-3}$  the  CI 2-1  line
 remains very  subthermally excited,  yielding $\rm R^{(ci)}_{21/10}$$\sim
 $1--1.8 (LTE values for  $\rm T_{kin}$=30-100\,K: $\sim $1.5-3.8), and
 only  for n$\sim$$\rm  (1-3)\times 10^{4}\,  cm^{-3}$ the  line ratio
 starts approaching its LTE~values.

 Finally Figure~\ref{R21tdF}  indicates the reason behind  the curious
 clustering  of the  (CI line/LTE)-derived  gas temperatures  within a
 narrow  range  of $\sim  $20-30\,K  which  was attributed  to  weakly
 varying average molecular gas  temperatures across galaxy populations
 (Valentino  et  al.   2020).   It  is instead  simply  an  effect  of
 subthermal CI 2-1  line excitation (and a  population $''$pile$''$ up
 on  the  J=1  level,  $\rm E_1/k_B$$\sim$24\,K),  under  the  typical
 densities of  the molecular  gas in  the ISM  of galaxies  not weakly
 varying gas temperatures.  Should all  the observed CI line ratios be
 interpreted using  solely the  LTE curve  (yellow), the  deduced $\rm
 T_k(CI, LTE)$  would be contained  within the $\sim  $20-30\,K range,
 while   for   numerous   Main  Sequence   galaxies   such   LTE-based
 interpretation would suggest average gas temperatures as low as $\sim
 $15\,K,  typical  for  the  ISM  of  SF-quiescent  not  of  SF-active
 galaxies.   On  the  other  hand, for  non-LTE  excitation  and  e.g.
 n=$10^3$\,cm$^{-3}$, the CI line ratio stays within a narrow range of
 values  ($\sim   $1-1.8)  even  as  $\rm   T_k$=(30-100)\,K  (a  more
 representative range  for the ISM  of galaxies  with a wide  range of
 SFRs).  Should the corresponding CI line ratios be interpreted in the
 LTE regime,  it would yield  $\rm T_k(CI, LTE)$$\sim $30)\,K  for the
 SF-extreme group.

 \section{Discussion}
\label{DIS}

\begin{figure*}
  \includegraphics[width=0.95\textwidth]{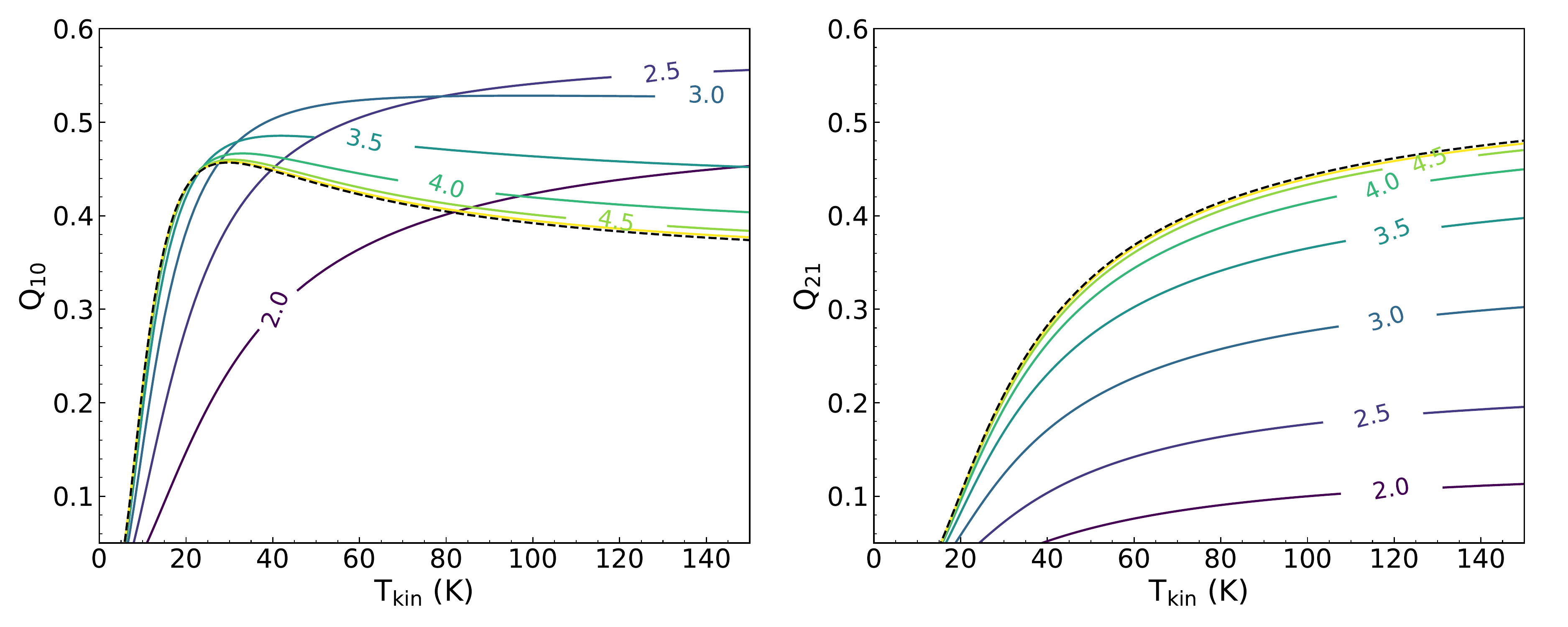}
\caption{\label{QF} The $\rm Q_{10}(n,\tk)$ (left) and the $\rm
  Q_{21}(n,\tk)$ (right) factors, for a range of gas densities and
  temperatures expected for the ISM of galaxies. Iso-density curves
  are labelled with log (n) /\cc, the LTE curve is shown as a black
  dashed line. It is note-worthy that for $\rm Q_{10}(n,\tk)$, the
  lower density curves lie above the LTE values for \tk>20--30~K
  (suprathermal) a behaviour explained in section~\ref{DIS}. 
   Thuse using the $\rm Q_{10}$(LTE) value at $\tk\sim$30~K will produce a {\em slight systematic
    overestimate} of the \CI\ mass (and hence \mol mass) when using
  CI(1--0) as a \mol tracer. For higher \tk\ this effect becomes more
  pronounced but is still $<35$\%.  Thus it does not
  pose a serious problem in the use of the CI(1--0) as a \mol
  tracer (unlike the adopted value of the CR-controlled [C/H$_2$] abundance).
   For the CI(2--1) line, the calibration as a tracer of \mol
  is far less straight-forward due to the large variation in $\rm Q_{21}$
  (0.05--0.45) at a given \tk, once we accept that the system is in
  non-LTE, while in the ISM of some galaxies  this line can be very faint
  because its  low sub-thermal excitation.}
\end{figure*}

The  subthermal  excitation  of  the   CI  lines  has  been  predicted
theoretically by Glover et al. 2015 using turbulent cloud simulations,
which also found the CI 1-0 line emission to be a good $\rm H_2$ cloud
structure   and  mass   tracer,   albeit  for   rather  small   clouds
($10^4$\,M$_{\odot}$) rather  than the typical Giant  Molecular Clouds
in  galaxies  ($\sim  $$10^5$-$10^6$\,M$_{\odot}$).  Our  study  lends
strong observational support to this view, which will impact molecular
gas  mass estimates  based on  the CI  lines, via  the so-called  $\rm
Q_{10}$=$\rm N_1/N_{tot}$  and $\rm Q_{21}$=$\rm  N_2/N_{tot}$ factors
which now become dependant on both the average density and temperature
of  the molecular  gas.  In  Figure~\ref{QF}  we show  the $\rm  Q_{10,21}(n,
T_{k})$ functions  obtained from  available analytic  expressions.  As
expected  the largest  sensitivity is  shown by  $\rm Q_{21}(n,  T_k)$
which varies  as $\sim  $0.05-0.45 for $\rm  T_{kin}$=(20--100)\,K and
n=($10^2$--$10^5$)\,cm$^{-3}$,  where it  remains $\rm  T_{ex}(2,1)\la
T_{kin}$ (i.e.  $\rm Q_{21}\la Q_{21}(LTE)$).

There is markedly less variation of the $\rm Q_{10}$ factor which, for
the  same  $\rm   (n,  T_k)$  range,  is:   $\sim  $0.3--0.5.   Perhaps
surpisingly, for n$\geq $$10^3$\,cm$^{-3}$,  its non-LTE values slightly
exceed     the     LTE     ones     (i.e.      $\rm     T_{ex}(1,0)\ga
T_{kin}$$\Leftrightarrow$$\rm   Q_{10}\ga   Q_{10}(LTE)$)  when   $\rm
T_{k}\ga 30$\,K.   Such slight superthermal excitation  occurs because
of   a  $''$pile$''$   up  on   the  J=1   level  induced   by  a   faster
J=2$\rightarrow$1  than  J=1$\rightarrow$0  spontaneous  de-excitation
(for the  CI lines:  $\rm A_{21}/A_{10}$$\sim $3.38)  once temperatures
are  high enough  for collisions  to  start populating  the J=2  level
(radiative  and  collisional  de-excitation  of  J=2$\rightarrow$0  is
negligible).  A similar effect can occur  for the CO J=2--1 and J=1--0
transitions as noticed  first by Goldsmith 1972 and  described in more
detail by Leung \&  Liszt 1976 (see also de Jong  et al. 1975), though
mostly for  the $^{13}$CO isotopologue  as the high optical  depths of
the CO lines act to surpress such non-LTE~effects.


The stronger  variation of the $\rm  Q$ factors in the  non-LTE regime
will affect the CI-based molecular gas mass estimates, adding one more
source  of  uncertainty  (besides  that of  the  $\rm  X_{CI}=[C/H_2]$
abundance)  to  the  molecular  gas mass/line  luminosity  ratio  $\rm
M(H_2)/L_{CI}\propto (X_{CI}  Q_{ik})^{-1}$. From  Figure~\ref{QF} the
mean  value of  $\rm  Q_{10}$ for  most of  the  expected average  ISM
conditions: $\rm [n, T_{kin}]$=$\rm [(300-10^4)\,cm^{-3}, (25-80)\,K]$
is  $\langle  Q_{10}\rangle$=0.48  with  $\la  $16\%  variation  (99th
percentiles),  and adopting  it  is adequate  for  molecular gas  mass
estimates via the CI 1-0 line for most average ISM conditions.

For the CI 2-1 line on the other hand, care must be taken to determine
the appropriate $\rm  Q_{21}$ factor if this line  (when detected) is to
be used  to estimate the  underlying molecular  gas mass. This  can be
performed  statistically per  galaxy  population,  in conjuction  with
other molecular  gas mass tracers  (dust, CO, $^{13}$CO)  (Dunne 2019;
Dunne et  al.  2021). Alternatively,  joint CI,  CO SLED and  dust SED
models  are a  powerful  way  to constrain  the  average ISM  thermal,
dynamical and  even chemical states  of the ISM in  invidual galaxies,
with  total $\rm  H_{2}$ gas  mass  estimates a  mere byproduct  (e.g.
Papadopoulos et al.  2014; Harrington  et al.  2021).  Future modeling
efforts  should incorporate  the analytical  3-level CI  line emission
model,  {\it without  any  assumptions about  the underlying  velocity
  field  (and its  role on  the line  formation mechanism)}.   This is
bound  to yield  more robust  models that  are less  sensitive to  the
well-known n-$\rm T_{k}$  CO SLED model degeneracies.   The latter are
due  to  the  significant  CO  line optical  depths,  and  the  common
expression  used  to incorporate  the  Large  Velocity Gradient  (LVG)
assumption into  the formula  for the  line escape  probability $\beta
_{ik}=f(n, T_{k},  dV/dR)$ for CO  line emission (which  is unecessary
for the CI lines).

The aforementioned promise of  an analytical model of the 3-level
  CI  line emission  incorporated  in joint  molecular  SLED and  dust
  emission SED  fits cannot be  fully realized without  including some
  new   astrochemistry,   the   one  regulating   the   average   $\rm
  R_{C}$=[C/H$_2$] abundance in the  ISM (entering the expressions for
  all relative  C and CO  line strengths).   It is expected  that $\rm
  R_{C}$=$\rm  f(n,  \langle  U_{CR}\rangle, Z)$,  with  astrochemical
  modelling  now mature  enough  to provide  such  functions, even  if
  empirical ones (see Bisbas et al.   2015 for such a function for the
  CO/H$_2$ abundance).   With adequate  line and continuum  data, such
  astrochemically-informed joint CO/CI SLED and dust emission SED fits
  hold  the additional  promise  of tying  an important  extragalactic
  parameter,  the average  SFR  density of  galaxies $\rm  \rho_{SFR}$
  ($\rm  \propto \langle  U_{CR}\rangle$), to  ISM line  and continuum
  observables.  The  latter may  not sound  much of  a feat  until one
  realizes: a)  the difficulty of practically  constraining a quantity
  like $\rm \rho _{SFR}$, especially in deeply dust-enshrouded mergers
  and the clumpy  SF disks typical at high redshifts,  and b) the role
  of $\rm \rho _{SFR}$ in setting the SF initial conditions in the ISM
  of  galaxies, and  perhaps  also the  stellar  IMF (Papadopoulos  et
  al. 2011).


\subsection{Molecular gas reservoirs outside SF galaxies:
  the CI line ratio versus $\rm T_{dust}$ as an indicator?}
\label{windCG}

A few of the most extreme starbursts in our sample (SFR$\ga $500\,$\rm
M_{\odot}$\,$\rm yr^{-1}$) have  such low global CI  line ratios ($\la
$1), that  can only be accommodated  by very low gas  densities n$\sim
$($10^2$--$3\times   10^2$)\,cm$^{-3}$   (Figure~\ref{R21tdF},   right
panel).  This is because their high $\rm T_{kin}$ values (deduced from
the  high $\rm  T_{dust}$ of  their global  SEDs), place  them to  the
low-n/high-T part  of Figure~\ref{R21tdF}. Such densities  are near to
the lowest ones found in Galactic  GMCs, and $'$hover$'$ near the $\rm
HI$$\rightarrow  $$\rm  H_2$  phase   transition  limit  expected  for
Galactic FUV fields ($\rm G_{\circ}$$\sim $1) in metal-rich ISM (Z=1).
The latter sets an absolute lower limit for gas densities if $\rm H_2$
is  to form  appreciably  in  a given  FUV  radiation and  metallicity
environment (e.g.  Papadopoulos,  Thi \& Viti 2002), a  limit that can
only  be higher  in  the  much more  FUV-intense  ISM environments  of
extreme starbursts.   Finally it is  unexpected to have such  very low
density gas dominate  the global CI line emission of  some of the most
extreme  starbursts,  where  large  mass  fractions  of  high  density
($\sim $$10^4$\,cm$^{-3}$) molecular gas  are expected (Gao \& Solomon
2004). We speculate that the  low-density gas component implied by the
low global  CI line ratios in  some of the most  extreme starbursts of
our  sample  does  not  reside inside  those  galaxies,  but  outside,
associated  with   powerful  molecular   gas  winds,   and/or  massive
reservoirs of Circumgalactic  (CG) gas and dust.  Such reservoirs have
been  found around  starbursts and  AGN  (Cicone et  al.  2018,  2021)
results of  powerful SF and/or  AGN feedback  on the ISM  of galaxies.
Global  CI line  emission in  such objects  may thus  have substantial
contributions from such  CG and/or wind-related $\rm H_2$  gas down to
very low densities\footnote{Sensitive CI  1-0 ALMA imaging has already
  revealed a massive outflow $\rm  H_2$ gas reservoir around the local
  starburst NGC\,6240 (Cicone et al. 2018)}.

In summary  the observed very low CI(2-1)/(1-0)  ratios and high
   $\rm T_{dust}$  values are  used (under the  1-component assumption
   for CI  line and dust continuum  emission) to deduce $\rm  H_2$ gas
   densities  that  are   far  too  low  to  belong   to  typical  GMC
   structures. Moreover this is obtained  for some of the most extreme
   starbursts in the  sample whose highly turbulent  ISM, if anything,
   is expected to have most of the $\rm H_2$ mass of the resident GMCs
   in  the  high density  domain  ($\rm  >10^3\, cm^{-3}$).   We  thus
   consider it more likely that the  implied low density $\rm H_2$ gas
   belongs  to  a  massive  molecular  wind and/or  CG  gas  and  dust
   component outside these starbursts.

Of course  in practice our  global CI  ratios and dust  SEDs encompass
{\it both}  the starburst ISM and  the wind/CG $\rm H_2$  gas and dust
reservoir, each with its own  distinct average $\alpha >1$ values, the
modest ones  associated with  the starburst ISM,  and the  higher ones
with the wind and/or CG $\rm  H_2 $gas and dust reservoirs outside the
galaxies.   In the  latter  case  the dust  (far  from the  SF-powered
FUV-fields inside the starburst) will be cold, while the $\rm H_2$ gas
will continue  to be  heated by  starburst-produced CRs  (that readily
convect out with the molecular  winds) and by shocks, yielding $\alpha
$  values higher  than those  in the  starburst ISM.   In this  simple
2-component variation of the 1-component framework used in this study,
the 1-component "composite" $\alpha $  will be skewed towards the high
$\langle \alpha \rangle $=2.5 values found by Harrington et al.  2021,
because  of a  massive wind/CG  $\rm H_2$  and dust  component present
outside their extreme starbursts. However the relative gas mass ratios
between these  two components cannot be  confidently constrained using
global SLEDs  and dust  SEDs like  the ones used  here, but  only with
spatially resolved CI and low-J CO line imaging observations {\it that
  remain     sensitive     to     large    scale     molecular     gas
  distributions}\footnote{Dust emission may not be  a good bet for the
  CG/wind component  as the  dust outside galaxies  is expected  to be
  very cold ($<15$\,K).}

Finally the global CI line ratios (along with well-sampled global dust
SEDs  for  constraining $\langle  \rm  T_{dust}\rangle$)  may be  good
indicators of galaxies with  possible massive, low-density, reservoirs
of CG  and/or molecular  gas winds  for reasons  that go  beyond their
excitation sensitivity to  the presence of very low  density $\rm H_2$
gas  that is  thermally decoupled  from  its concomitant  dust. It  is
because in  the CR-irradiation environments  of CG and  molecular wind
material very low-density gas  $\sim $(50-200)\,cm$^{-3}$ will be very
CO poor (Papadopoulos  et al. 2018) and thus very  faint even in low-J
CO line emission. Moreover, even  if CO could somehow form efficiently
at such low densities while being subjected to the CR energy densities
of the wind/CG environments, the high-J CO lines necessary for probing
high-$\rm T_{kin}$  gas (and  thus for  identifying molecular  gas and
dust with extreme gas-dust thermal decoupling) unfortunately also have
considerable  critical  densities.   Indicatively CO  4-3,  with  $\rm
E_{4}/k_B$$\sim $55\,K (comparable to the $\rm E_2/k_B$$\sim $62\,K of
the    CI    2-1    line,   has    $\rm    n_{crit}$$\sim    $$2\times
10^4$\,cm$^{-3}$$\sim $20$\times$$\rm n_{crit}$(CI,  2-1). Such high-J
CO lines  will thus  be faint for  warm {\it and}  low density  gas in
molecular  gas winds  and CG  reservoirs, and  imaging their  emission
cannot reveal their true extent and mass around galaxies. Their global
(unresolved)  emission on  the other  hand  will be  dominated by  the
starburst ISM rather than by the  feebly emitting (in high-J CO lines)
molecular wind and CG gas reservoirs.

Of course, once a candidate starburst/AGN is selected for the possible
presence  of massive  $\rm H_2$  wind and/or  CG reservoirs  using the
global CI line/dust emission criterion,  deep CO line imaging of low-J
CO     lines\footnote{J=1-0,    2-1,     3-2,     so    that     their
  (radiative-trapping)-reduced   critical    densities   remain   $\la
  10^{3}\,cm^{-3}$ and do not become overly biased against low-density
  gas.}  remains  valuable for  independently identifying  the CO-rich
parts of  the wind and/or  CG molecular gas reservoirs,  and outlining
their  distribution and  kinematics (see  Cicone  et al.   2021 for  a
recent such result).  Sensitive  CI line interferometric imaging, that
includes short spacing and total  power measurements (so that extended
emission  from the  CG  and/or  molecular gas  wind  reservoir is  not
resolved out), can then uncover any  CO-poor low density $\rm H_2$ gas
along with the CO-rich~one.
 
 In the case  of the high frequency CI lines  such observations can be
 challenging, carrying the very real possibility of resolving out much
 of  the   extended  CG  and   molecular  gas  wind   emission.   This
 necessitates the  most compact ALMA/ACA configurations  aided by deep
 observations in total  power (TP) mode.  In that  regard, the planned
 50-m    class    AtLAST    submm    telescope    in    the    Atacama
 Desert\footnote{https://www.atlast.uio.no/}, with a combination
   of suitable  heterodyne receivers and  bolometers, can be  a search
   $''$machine$''$ for massive, thermally-decoupled, molecular gas and
   dust reservoirs  around distant SF  galaxies and QSOs,  outcomes of
   strong AGN and/or SF feedback  processes across the Universe.  When
   it comes to  imaging such reservoirs around  the candidate objects,
   spectral lines of low-J CO and  CI (for H$_2$), and CII (for H$_2$,
   HI,  and HII),  may  be the  sole capable  tracers  since the  dust
   continuum may  be very faint  because of the low  dust temperatures
   expected far from the  starburst/AGN\footnote{At high redshifts the
     difficulty of imaging such reservoirs, no matter how massive, via
     their dust continuum, is further compounded by their immersion in
     a rising CMB, inducing very low dust-CMB emission contrast (Zhang
     et al. 2016)}, and the  expected high gas-dust thermal decoupling
   at low gas densities.



\section{Conclusions}

Guided by an ever increasing body  of CI 1-0, 2-1 line luminosity data
for  galaxies  across  cosmic  epoch, and  their  use  as  alternative
molecular gas temperature and  mass distribution tracers, we assembled
a large  dataset of 106 galaxies  with available total CI  2-1 and 1-0
line luminosities  in order  to examine  their excitation  levels over
galaxy-sized  molecular   gas  mass  scales.   Using   the  analytical
solutions of the 3-level system responsible for the CI lines, the dust
temperatures obtained from global dust emission SEDs available for our
sample, along  with the  expected gas-dust thermal  decoupling factors
for FUV/CR-irradiated ISM, we reached the following conclusions:

\begin{itemize}

\item The CI lines are subthermally (non-LTE) excited over the bulk of
  the ISM  of galaxies, and  as such their ratio  cannot be used  as a
  probe of average molecular gas temperatures.

\item The recently  reported small variation of  the average molecular
  gas temperatures  in galaxies, deduced  using the CI line  ratio and
  the  assumption of  LTE,  is  an artifact  of  the  non-LTE CI  line
  excitation  conditions,  and does  not  reflect  actual  average
  molecular gas temperatures.
  
\item  The non-LTE  line  excitation impacts  the  molecular gas  mass
  estimates based on the CI lines.   For the CI 1-0 line, adopting the
  mean  non-LTE excitation  factor  $\langle  Q_{10}\rangle =0.48$  is
  adequate,  as  it  varies  little ($\sim  $16\%)  over  typical  ISM
  conditions variations.  For the CI 2-1 line this uncertainty is much
  larger with the corresponding $\rm Q_{21}$ factor varying over $\sim
  $0.05--0.45.

\item The analytical  non-LTE expression of the CI  2-1/1-0 line ratio
  $\rm R^{(ci)} _{21/10}(n, T_k)$ can be  used in joint models with CO
  lines and dust  emission SEDs without adopting  the assumptions used
  for CO line radiative transfer models (i.e.  Large Velocity Gradient
  velocity fields  and local escape probability  expressions depending
  on them and the assumed cloud  geometry). This will add robstness in
  future joint CI/CO SLED and dust SED models for the ISM of galaxies.

\item Some  of the lowest  ($\la 1$) global CI(2-1)/(1-0)  line ratios
  and high average $\rm T_{dust}$  values (and thus $\rm T_{kin}$) are
  found  in  some  of  the  most extreme  starbursts  of  our  sample,
  indicating large amounts of warm but very low density molecular.  We
  conjecture  that  this  molecular   gas  reservoir  is  outside  the
  starburst/AGN, part  of a  powerful wind and/or  Circumgalactic (CG)
  material, and propose this combination  of low CI(2-1)/(1-0 and high
  $\rm T_{dust}$ values as a criterion to locate them.

 \end{itemize}


\section*{Data availability}
No new data were generated or analysed in support of this research.
 
\section*{Acknowledgements}

We would like to thank the anonymous referee for comments that allowed
us to  clarify issues regarding  CI line  optical depths, as  well the
astrophysics behind  the gas-dust thermal decoupling  and the possible
values of  the $\alpha  $ parameter in  galaxies. The  latter directly
informed  our now expanded discussion in section 3.1.

Finally we thank  our few remaining friends for keeping us old sailors 
onboard the  pirate ship of  Discovery, and not tossing us overboard, to
the placid Sargasso  Seas of irrelevance and Oblivion.  PPP would like
to thank Seren for being such an accommodating little angel during his
visit in  Cardiff when he  experimented endlessly in the  kitchen, and
her parents survived...




\bsp	
\label{lastpage}
\end{document}